\documentclass[11pt]{article}
\usepackage{moriond,epsfig}

\def\be{\begin{equation}}
\def\ee{\end{equation}}
\def\bea{\begin{eqnarray}}
\def\eea{\end{eqnarray}}

\begin{document}
\title{EFFICIENT QUANTUM COMPUTING INSENSITIVE TO PHASE ERRORS}

\author{\underline{B. GEORGEOT},
D. L. SHEPELYANSKY}

\address{Laboratoire de Physique Quantique, UMR 5626 du CNRS, 
Universit\'e Paul Sabatier, F-31062 Toulouse Cedex 4, France}

\maketitle\abstracts{
We show that certain computational
algorithms can be simulated on a quantum computer with exponential efficiency
and be insensitive to phase errors.  
 Our explicit algorithm simulates accurately
the classical chaotic dynamics for exponentially many orbits 
even when the quantum fidelity drops to zero.  Such phase-insensitive
algorithms open new possibilities for computation 
on realistic quantum computers.}

\section{Introduction }

The problem of quantum computation has attracted recently a great
deal of attention \cite{divincenzo,josza,steane1}.  This interest
stems from the fact that the massive parallelism 
permitted by quantum mechanics
enables to reach exponential efficiency of computation in certain quantum
algorithms.  The most famous example is the Shor algorithm which
allows to factor large numbers exponentially faster than any known classical 
algorithm \cite{shor1}.  Recently, other types of exponentially efficient 
algorithms has been developed for the simulation of various physical systems
\cite{lloyd,qckr,cat}.  The exponential gain in computation is related to the 
exponentially large Hilbert space of the quantum computer which is composed
of multi-qubit states
operating in parallel (each qubit is a two-level quantum system). 
 Usually the algorithms  are  constructed for ideal
quantum computers operating free of noise and imperfections.  
In reality, any physical realization of such a computer involves a certain
level of imperfections, noise in gate operations and decoherence.  First
investigations showed that quantum computation can tolerate a sufficiently low 
level of errors \cite{zoller,paz}.  More recently, it has been found
that quantum computation is tolerant to quantum errors when simulating 
classical chaotic dynamics for which classical computer errors grow 
exponentially
with time \cite{cat}. However in general the quantum errors grow 
with the number of gate operations and any realistic quantum computer
is faced with this problem.  To deal with this problem of fault-tolerant 
computation, quantum error-correcting codes were recently developed
\cite{shor2,steane2,preskill}.  They allow to reduce the level of errors
in a systematic way, but require the introduction of many redundant qubits
and additional gates, which significantly complicates the computational
process.  The complexity of these codes depends strongly on the type of errors 
they should correct.  Indeed, while simpler classical codes need to correct only
bit errors, the quantum ones should in addition simultaneously correct 
the quantum phase errors.  The quantum phase errors seem to be of
primary importance since the massive parallelism of quantum computing is
related to entanglement in the Hilbert space which is directly based on 
phase coherence.  Therefore, according to this common lore it seems
impossible to perform efficient and accurate quantum computations in
presence of uncontrolled strong phase errors.  In this paper we show
on an explicit example that it is not always the case, and that there
are algorithms insensitive to phase errors 
which perform accurate and efficient computation.  Our example is based 
on the simulation of classical chaotic dynamics, which is very hard to 
simulate accurately on classical computers since this dynamics is unstable
and round-off errors grow
exponentially with time.  In spite of this, our quantum algorithm, including
measurement,  remains 
insensitive to phase errors for arbitrary time.

\section{The Arnold-Schr\"odinger cat quantum algorithm}

To illustrate this phenomenon, we choose an algorithm which simulates
the classical chaotic dynamics of the well-known Arnold cat map 
\cite{arnold,lieberman}.  It was
recently shown \cite{cat} that quantum computers can simulate this dynamics
with exponential efficiency.  In addition it was shown that
a small level of quantum errors in the gate operations of order $\epsilon$
allows to simulate accurately
this map on times of order $O(1/\epsilon^2)$.  Thus quantum
computers can face classical exponential instability and chaos.

The dynamics of the map we are studying is given by:
\begin{equation}
\label{catmap}
\bar{y}=y+x \; \mbox{(mod} \;\mbox{1)}\;\;, \;\; \bar{x}=y+2x \;\mbox{(mod} 
\;\mbox{1)}\;,
\end{equation}
where bars denote the new values of the variables after one iteration.
 This is an area-preserving map, in which $x$ can be considered as 
the space variable and $y$ as the momentum. A discretized version on 
a $N \times N$ square lattice is also described by this map. In \cite{cat}, 
a quantum algorithm called Arnold-Schr\"odinger cat map was introduced, and it
was shown to simulate this dynamics on the lattice with exponential efficiency.
In this paper, we modify this algorithm in such a way that exponential
efficiency is preserved and in addition it becomes insensitive to phase errors.
This is obtained by the introduction of a new measurement procedure.

The quantum algorithm introduced in \cite{cat}
simulates the discrete classical dynamics given
by (\ref{catmap})
and operates with $3n_q -1$ qubits.  These qubits are organized in three 
registers. Two of them describe the classical phase space 
with $N^2$ points and $N=2^{n_q}$.  
The third register with $n_q -1$ qubits is used as 
workspace.  In this way, an initial classical phase space density 
can be represented by a
quantum state $\sum_{i,j} a_{ij} |x_i> |y_j>|0>$, with 
$x_i=i/N$, $i=0,...,N-1$ and $y_j=j/N$, $j=0,...,N-1$, written in binary 
representation, and we choose initially $a_{ij}=0$ or $1/\sqrt{N_d}$ where
$N_d$ is the number of points in the classical distribution.
Then, iterations of the map (\ref{catmap}) are performed with
the help of additions of integers modulo (N) (modular additions).
The quantum algorithm we use for this operation is similar to the one described
in \cite{barrenco} (see also \cite{cat}).  First we compute all the carries
of the addition, using two Toffoli gates and one controlled-not 
(CNOT) gate per qubit.  Then we perform the addition starting from the last qubit
and erasing the carries by running the inverse of the preceding step.  This
needs two CNOT gates per qubit addition and the same gates as above to 
erase the carries.  The result is taken modulo (N) by eliminating
the last carry.  After these operations, the amplitudes $|a_{ij}|$ describe
the classical phase space distribution function after iteration of 
($\ref{catmap}$). In total, one needs $16n_q-22$ Toffoli and CNOT gates
per map iteration.  On the contrary, a classical computer will require
$O(2^{2n_q})$ operations per iteration for $N_d= O(N^2)$ orbits.

It is important to stress that during the whole process the classical
distribution function is determined only by the probabilities $|a_{ij}|^2$ of
the quantum computer wavefunction expanded on the Hilbert space basis of
register states $|x_i> |y_j>$ (after each map iteration the third register
is always in the state $|0>$).  Thus, the information about the classical
distribution function is stored in these probabilities, and is not sensitive
to the relative quantum phases of $a_{ij}$.  This suggests that the phase
errors accumulated during gate operations will not affect the quantum
computer simulation of this classical chaotic dynamics.  

\section{Effect of phase errors}

To study the effects of quantum errors, one usually uses 
the fidelity \cite{paz},
defined as: $f(t)=$ 

\noindent$|<\psi_{\epsilon}(t)|\psi_0 (t)>|^2 = 
|\sum_{i,j} a_{ij}^{(\epsilon)}(t) a_{ij}^{*(0)}(t)|^2$.
Here $|\psi_0 (t)>=\sum_{i,j} a_{ij}^{(0)}(t) |x_i> |y_j>$ 
is the quantum state after $t$ perfect iterations, while
$|\psi_{\epsilon}(t)>=\sum_{i,j} a_{ij}^{(\epsilon)}(t) |x_i> |y_j>$ 
is the quantum state after $t$ imperfect iterations.  Obviously, this quantity
is very sensitive to the relative phases of $a_{ij}$.  Since the classical
phase space density is not sensitive to these phases, we introduce another
characteristic which is related only to the amplitudes $|a_{ij}|$.  We call it
{\em faithfulness} and define it by: 
$f_{a}(t)=(\sum_{i,j} |a_{ij}^{(\epsilon)}(t) a_{ij}^{(0)}(t)|)^2$. 
This quantity can be considered as a generalization of the usual fidelity. 
As well as $f(t)$, the faithfulness $f_a(t)$ is always $\leq 1$, 
and it determines the deviation from the exact amplitudes (the value $1$ 
is reached only for $|a_{ij}^{(\epsilon)}(t)|=|a_{ij}^{(0)}(t)|$ for all $i,j$).
In addition, one has always $f_a(t) \geq f(t)$.
Contrary to the usual fidelity, the faithfulness measures only amplitude errors, 
being insensitive to the quantum phases.  We note that its definition is related
to the preferential basis chosen initially in the Hilbert space.  
 
To study the dependence of fidelity and faithfulness on phase errors, 
the nondiagonal part of each Toffoli and CNOT gate used in the algorithm 
was multiplied by 
a diagonal matrix with elements
$\exp(i\theta_m)$, with random phases $\theta_m$ homogeneously distributed 
in $[-\epsilon_{\phi}, \epsilon_{\phi}]$.  The results are shown in Fig.~1 (Left).
Here the initial state was chosen in the form of a cat's smile (see Fig.~1
of \cite{cat} and the coarse-grained version in Fig.~2). In the presence
of phase errors only, the fidelity $f$ decreases with the number 
of map iterations
and drops almost to zero for sufficiently strong phase noise. At the same time,
the faithfulness $f_a$ is not affected even by the maximal possible phase
noise.  We also checked that $f_a=1$ is not affected 
if each $a_{ij}$ is multiplied after each gate by a random phase 
$\exp(i\theta_m)$
with $\theta_m$ distributed in $[-\pi, \pi]$, although in this case the 
fidelity is almost zero after one map iteration.  

\begin{figure}[ht]
\begin{center}
\psfig{figure=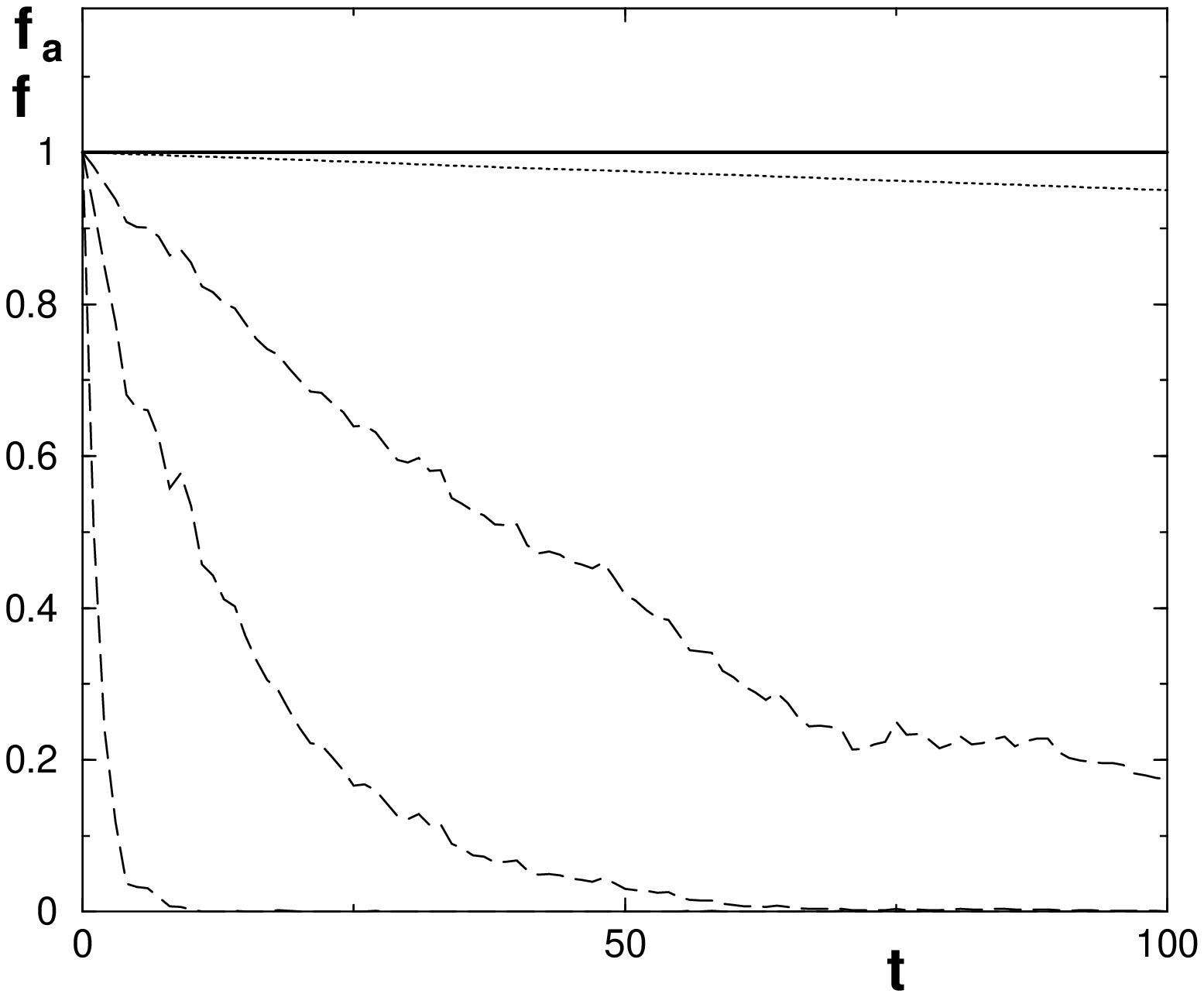,height=5.5cm,width=7.5cm}
\psfig{figure=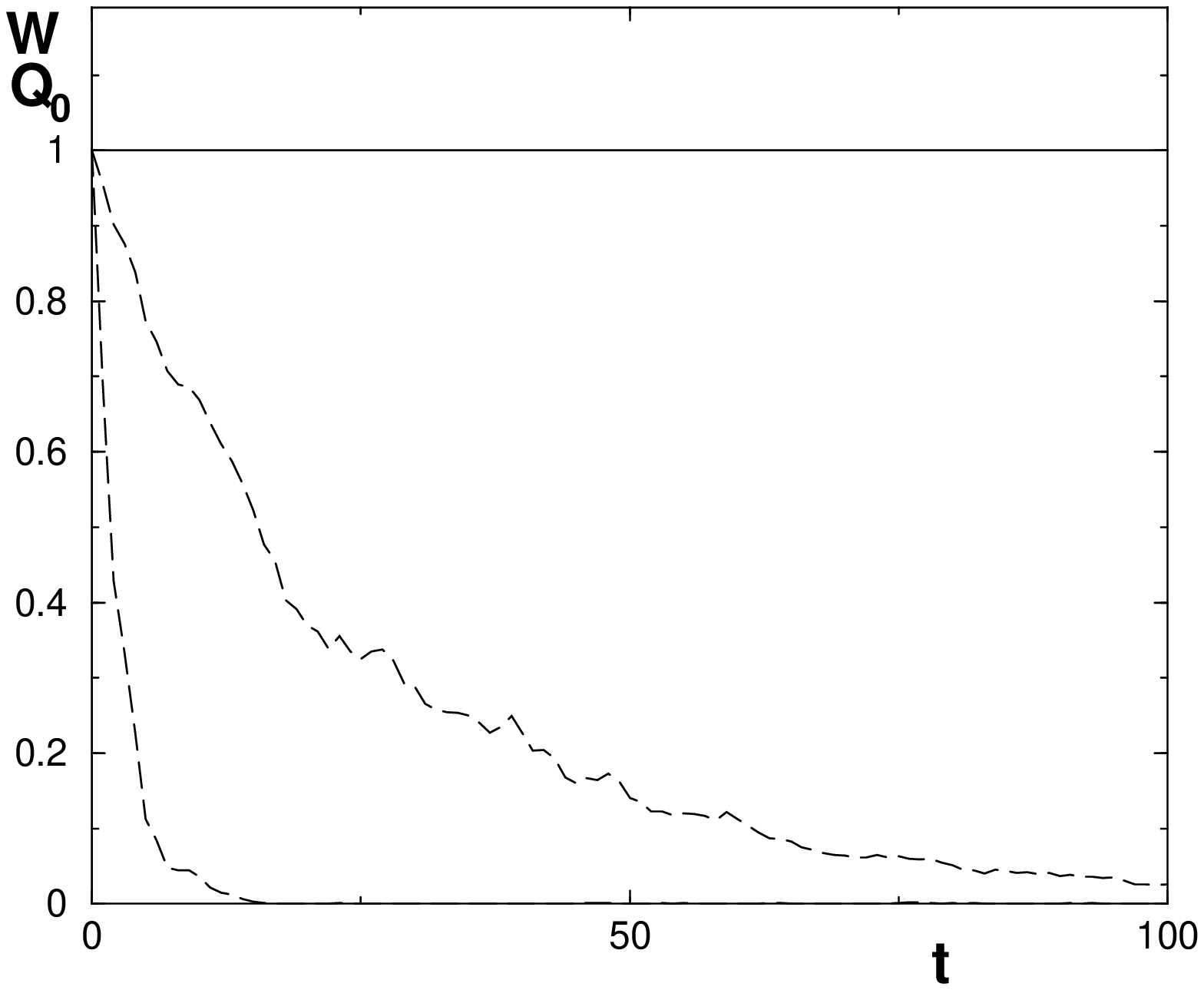,height=5.5cm,width=7.5cm}
\end{center}
\caption{
LEFT: Quantum fidelity $f$ of Arnold-Schr\"odinger cat as a function of
 time $t$ for quantum phase errors
of strength $\epsilon_{\phi} = 0.05, 0.1, 0.3$
(dashed curves from top to bottom ).
Faithfulness $f_a$ is shown for the maximal phase errors with 
$\epsilon_{\phi}=\pi$ 
 (full line). The dotted line shows $f_a(t)$ when in addition to maximal phase
errors there are small amplitudes errors of strength $\epsilon=0.01$.
Initial state is chosen in the form of a cat's smile on a 
$ 128 \times 128$ lattice (see text and Fig.~2),
and $n_q=7$. In total, $20$ qubits are used for the computation.
$\;\;\;\;\;\;\;\;\;\;\;\;\;\;\;\;\;\;\;\;\;\;\;\;\;\;\;$
RIGHT: Zero harmonic $Q_0$ of Arnold-Schr\"odinger cat normalized
by its value in absence of errors as a function of
 time $t$ for quantum phase errors
of strength $\epsilon_{\phi} = 0.07, 0.2$
(dashed curves from top to bottom ).
Full line shows the total probability $W$ 
inside one cell ($i_g', j_g'$) (see text)
normalized in the same way for the maximal phase errors with 
$\epsilon_{\phi}=\pi$.
Initial condition is chosen as for Left, $n_q=7$ and $n_g=5$.}
\label{fig1}
\end{figure}

To show the difference between phase and amplitude (bit) errors, we computed
the faithfulness in presence of a small unitary noise in the gates.  
For that, in addition to large phase errors,
for each gate the nondiagonal 
part was diagonalized, and each eigenvalue was multiplied by a random phase 
$\exp(i\eta)$, with $-\epsilon <\eta < \epsilon$.  This introduces both phase
and amplitude errors, and Fig.~1 (Left) shows that the faithfulness starts 
to drop
slowly with time.  Hence despite the presence of strong phase errors 
the faithfulness is sensitive only to the amplitude errors.

Thus the information stored in the amplitudes is not sensitive to phase
decoherence.  Still, one should find a way to retrieve a part 
of this exponentially large information.  Usually one performs a quantum
Fourier transform (QFT) and measures the maximal harmonics of the distribution,
as was suggested in \cite{cat}.  However, the QFT is extremely sensitive
to the quantum phases of $a_{ij}$, as is illustrated in Fig.~1 (Right).  
The zero
harmonic $Q_0(t)=\sum_{i,j} a_{ij}(t)/N =\sqrt{N_d}/N$ 
is time-independent in the absence of errors, but drops rapidly with $t$ if
phase errors are present.  To avoid the effects of phase errors, 
one can measure only the first $n_g$ qubits from the $n_q$ qubits present 
in the register $|x>$ and the same for the register $|y>$.  This procedure
introduces a coarse-graining of the phase space ($x,y$), with the number of
cells $N_g=2^{2n_g}$.  The result of such a measurement is determined by the
total probability inside each cell $W_{i_g j_g}=\sum_{<i,j>} |a_{ij}|^2$
where the summation is performed over all ($i,j$) inside the cell ($i_g, j_g$).
This probability is not sensitive to phase errors, and can be extracted
by a number of measurements which is polynomial in $N_g$.  Fig.~1 (Right)
 shows that 
the probability in a chosen cell ($i_g', j_g'$) is indeed insensitive to phase
errors.  We note that this coarse-grained probability is a very natural
quantity for the dynamical system under investigation.  One is not
interested in the exponential amount of information present in all $a_{ij}$
since one cannot even store it classically, and therefore it is better
to operate with coarse-grained characteristics as is usually done
in chaotic dynamical systems.  The number of cells $N_g$ can be kept 
constant while the number of iterated classical
orbits increases exponentially with $n_q$.  All these orbits are iterated 
accurately and with
exponential efficiency during quantum computation, and the constant number
of cells $N_g$ is used only to extract the essential information generated
by this chaotic dynamics.

\begin{figure}[ht]
\begin{center}
\psfig{figure=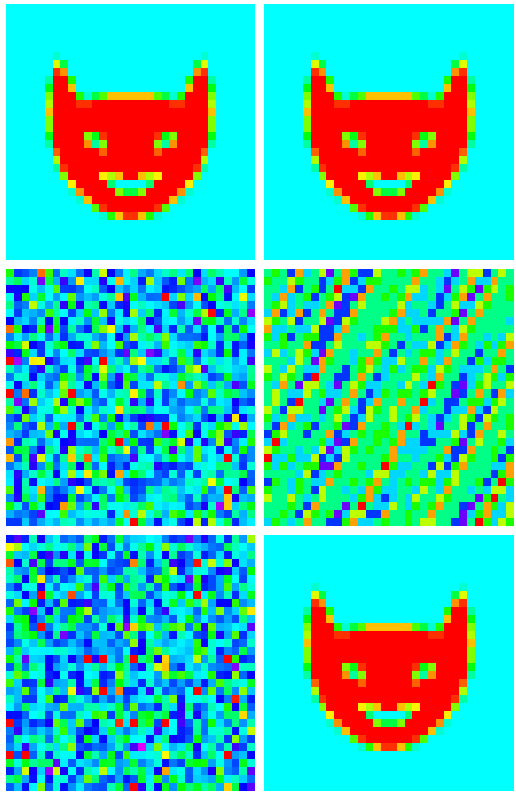,height=11.25cm,width=7.5cm}
\end{center}
\caption{Coarse-grained image of Arnold-Schr\"odinger cat 
measured through the probabilities $W_{i_g j_g}$ 
at different moments of time $t=0$(upper panels),
$t=50$ (middle panels), $t=100$ (lower panels).  
%
Color is proportional
to $W_{i_g j_g}$, from blue (minimum) to red (maximum).
Time inversion
is done at $t=50$. The quantum computation
is done with phase errors of amplitude $\epsilon_{\phi}=\pi$
for both columns.  In addition, for the left column the computation
includes also amplitude errors  of strength $\epsilon=0.3$. The
initial state is as in Fig.1, and $n_q=7$. The coarse-graining corresponds
to measuring the first five qubits ($n_g=5$) in the $|x>$ and $|y>$ registers.
}
\label{fig2}
\end{figure}

Another test of the effects induced by phase and amplitude errors can be
performed on the basis of time-inversion.  Indeed, the exact map (\ref{catmap})
is exactly time-invertible.  However, in presence of imperfections this 
reversibility can be destroyed.  In the r\'egime of classical chaos, the
classical round-off errors grow exponentially with time and destroys
time-reversibility in a logarithmically short time.  For quantum
simulations, it was shown in \cite{cat} that quantum errors grow only
polynomially with time.  Due to that, time-reversibility is preserved in 
quantum computation of (\ref{catmap}) for relatively small errors.  However,
it is naturally expected to be destroyed in the case of strong errors.  
Contrary to this expectation, Fig.2 (Right) shows that time-reversibility 
is exactly preserved in the presence of phase errors of maximal amplitude,
and the classical distribution is exactly reproduced for all $t$. 
In the right column, the difference between the two images at $t=0$ and
$t=100$ is on the level of classical computer precision.  On the contrary,
strong enough amplitude errors completely destroy time-reversibility,
as is shown on Fig.2 (Left).

\section{Discussion}

Thus, all the data clearly show that our quantum algorithm 
simulating (\ref{catmap}) is insensitive to phase errors.  This result can
be understood in the following way.  All nondiagonal parts of the
gates used in the algorithm are represented by the operator $\sigma^x$,
while the noncommuting part of phase errors is represented by
$\sigma^z$.  Of course, $\sigma^x$ and $\sigma^z$ do not commute.  However,
the action of $\sigma^x$, $\sigma^z \sigma^x$ and $\sigma^x \sigma^z$ on a
two-component spinor gives the same amplitudes of the components 
(with different relative phases).  Thus any quantity encoded in the amplitudes,
in our case the classical distribution function, remains invariant in presence
of $\sigma^z$ (phase) errors. On the contrary, it is 
sensitive to $\sigma^x$ (amplitude) errors.
 Another way of understanding
this insensitivity to phase errors is to remark that all used gates belong
to a very specific subgroup among unitary transformations of the Hilbert space,
that is the group of permutations of the basis formed by the states where each
qubit is polarized in the $z$ direction (each qubit is either $|0>$ or $|1>$).
The amplitudes in this basis are insensitive to phase errors if only such
transformations are present in the algorithm.  Indeed, any permutation can be 
written as a product of transpositions which exchange only two states.  
By the same argument as for $\sigma^x\sigma^z$ given above, 
such a transformation is immune to phase errors, and hence 
any permutation.  We stress again that phase errors do affect the final
state through the relative phases, but do not affect the measurement 
which gives the cell probabilities $W_{i_g j_g}$. 

The above mathematical argument explains the insensitivity
to phase errors.  In a more physical way, we can say 
that the map (\ref{catmap}) describes the classical dynamics of
the Arnold cat map, which naturally should not be sensitive to quantum phases.
Of course, one can imagine other quantum algorithms which will
simulate this classical dynamics using both phases and amplitudes of the wave
function, and therefore will be sensitive to phase errors.  However, on the
basis of the Arnold-Schr\"odinger cat algorithm discussed in this paper, 
we make the conjecture that classical Hamiltonian dynamics of generic
systems can always be 
simulated on a quantum computer in a way insensitive to phase errors.  Indeed,
for such a dynamics the classical information can be naturally encoded in the
amplitudes only \cite{standard}.  
It is rather likely that such a situation can appear in
quantum computations which are not connected with classical mechanics, 
for example probing the range of values of a function.

The implementation of such algorithms insensitive to phase errors
can be enormously simpler than in the case of other algorithms sensitive to 
quantum phases.  Indeed, the necessity to correct both phase and amplitude
errors significantly complicates quantum error-correcting codes
\cite{shor2,steane2,preskill}.  If only amplitude errors are to be corrected,
one can use much simpler codes close to the classical ones.  Also, in some
physical systems phase errors can be naturally much stronger than amplitude 
ones.  For example, recent studies of the emergence of quantum chaos
in a quantum computer \cite{nous} showed that for sufficiently
strong residual inter-qubit interaction, exponentially many states are mixed
and amplitude errors become enormously strong.  On the contrary, below the
quantum chaos border, amplitude errors are very small whereas phase errors
are still important.  In spite of that, a quantum computer in this r\'egime
can efficiently simulate algorithms of the type discussed here.  

We note that while the algorithm presented above is exponentially
faster than any  deterministic classical algorithm iterating the map 
(\ref{catmap}) nevertheless one can try to compete with 
it with the help of classical Monte Carlo simulation
with a polynomial number of trajectories. Such an approach
does not produce the exact density distribution
with an exponential number of orbits which is hidden in the
quantum wavefunction. However, the statistical accuracy of both methods
can be comparable since one makes a polynomial number of
measurements of the quantum final state.
At the same time one should keep in mind that
such a Monte Carlo simulation is based on the statistical assumption
that a polynomial number of  trajectories can correctly
describe the fine structure of classical phase space.
In contrast, the quantum simulation takes exactly into account the dynamics
on {\it all} scales. We also stress that without large phase errors
the QFT gives access to information which is unaccessible even
for classical Monte Carlo algorithms.

In conclusion, we have shown the existence of quantum algorithms which
can simulate efficiently certain computational problems and at the 
same time are insensitive to phase errors.  Our explicit example
is related to the simulation of classical 
motion and we make the conjecture that 
classical Hamiltonian dynamics can always be simulated in a way immune
to phase decoherence.  The existence of such efficient quantum algorithms 
insensitive to the relative phases shows that contrary to the common lore,
the massive parallelism of quantum computing is not necessarily related
to quantum interference.  Actually, quantum mechanics allows to follow 
in parallel exponentially many computational paths in a way insensitive
to phase decoherence.

{\em \section*{Acknowledgments}}
We thank C.~W.~J.~Beenakker for discussions and
the IDRIS in Orsay and the CalMiP in Toulouse for access to their
supercomputers. This work was supported in part 
by the EC RTN contract HPRN-CT-2000-0156.


\begin{thebibliography}{99}
\bibitem{standard}This is the case for the algorithm proposed in \cite{qckr} 
to simulate the classical Chirikov standard map.
\end{thebibliography}
\end{document}